\documentclass[12pt]{iopart}

\usepackage{graphicx}

\usepackage{iopams}
\usepackage{pslatex}
\usepackage{txfonts}

\newcommand{\sqrts}{\sqrt{s}}
\newcommand{\sqrtsnn}{\sqrt{s_{_{NN}}}}
\providecommand{\jpsi}{J/\psi}

\providecommand{\ups}{\Upsilon}

\providecommand{\mean}[1]{\ensuremath{\left<#1\right>}}
\providecommand{\qqbar}{Q\overline{Q}}

\providecommand{\dNdeta}{dN_{ch}/d\eta|_{\eta=0}}
\providecommand{\dNgdy}{dN_{g}/dy}

\providecommand{\hydjet}{{\sc hydjet}}

\def\mean#1{\ensuremath{\left<#1\right>}}

\providecommand{\qhat}{\mean{\hat{q}}}
\providecommand{\percms}{cm$^{-2}$s$^{-1}$}

\newcommand {\etg}         {\ensuremath{E^{\gamma}_{\mathrm{T}}}}
\newcommand {\etj}         {\ensuremath{E^{\rm jet}_{\mathrm{T}}}}

\providecommand{\PbPb}{Pb-Pb}
\providecommand{\gaga}{\gamma\,\gamma}

\providecommand{\gA}{\gamma\,A}
\providecommand{\gpb}{\gamma\,$Pb$}

\providecommand{\elel}{e^+e^-}
\providecommand{\mumu}{\mu^+\mu^-}

\begin{document}

\title[High-density QCD with CMS at the LHC]{High-density QCD with CMS at the LHC}
\author{David d'Enterria for the CMS collaboration}
\address{CERN, CH-1211 Geneva 23, Switzerland}

\begin{abstract}
The capabilities of the CMS experiment to explore the rich heavy-ion physics programme 
offered by the CERN Large Hadron Collider (LHC) are summarised. Various representative 
measurements in Pb-Pb collisions at $\sqrtsnn$~=~5.5 TeV are covered. These include 
``bulk'' observables -- charged hadron multiplicity, low-$p_{T}$ inclusive 
hadron spectra and elliptic flow -- which provide information on the collective properties 
of the system; as well as perturbative processes -- high-$p_{T}$ hadrons, 
jets, $\gamma$-jet and quarkonium production -- which yield ``tomographic'' information of the 
densest phases of the reaction.
\end{abstract}


%

\section*{Introduction}

\begin{figure}[!Hhtb]
\includegraphics[width=7.5cm,height=8.cm]{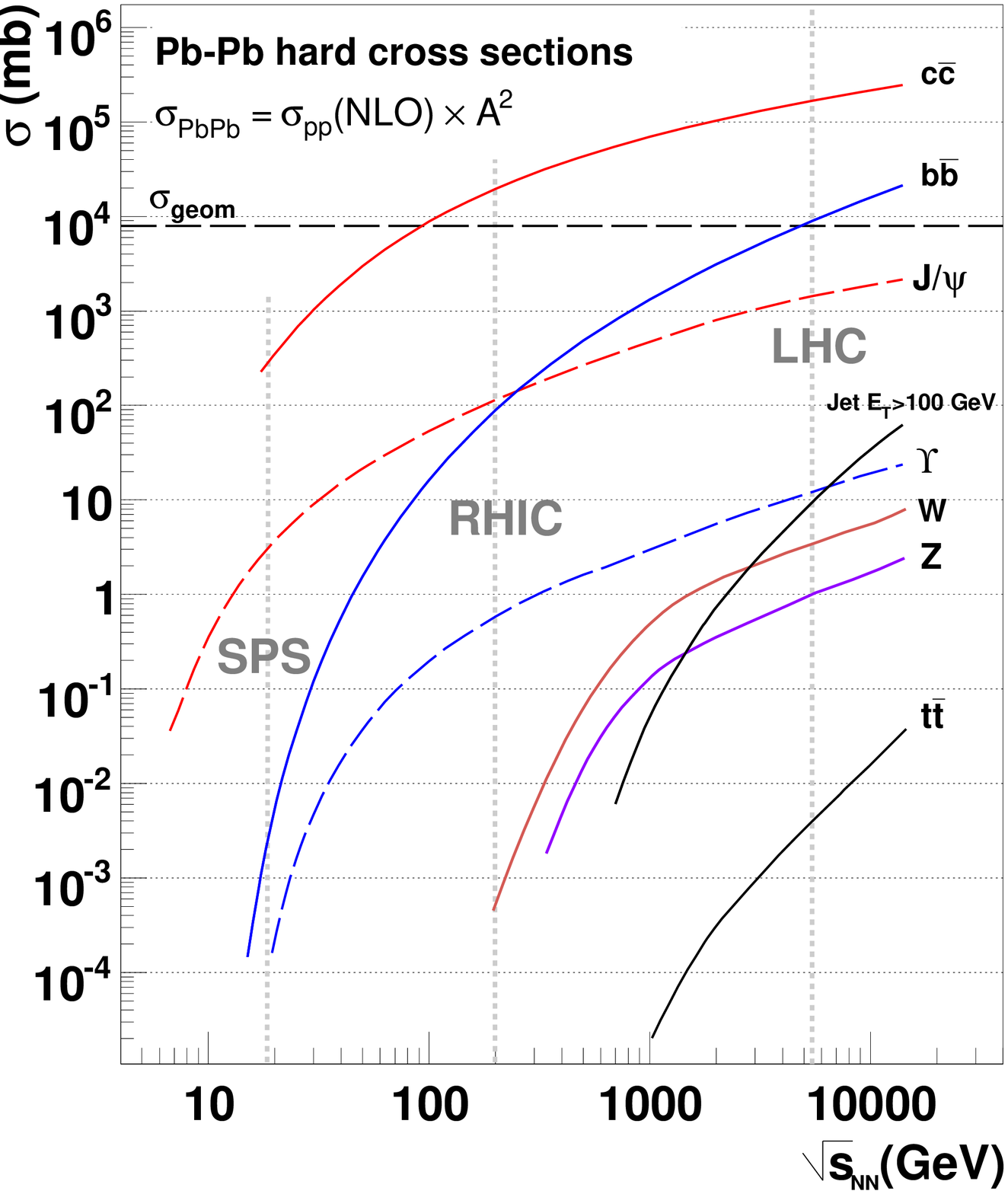}\hspace{0.4cm}
\includegraphics[width=7.5cm,height=7.95cm]{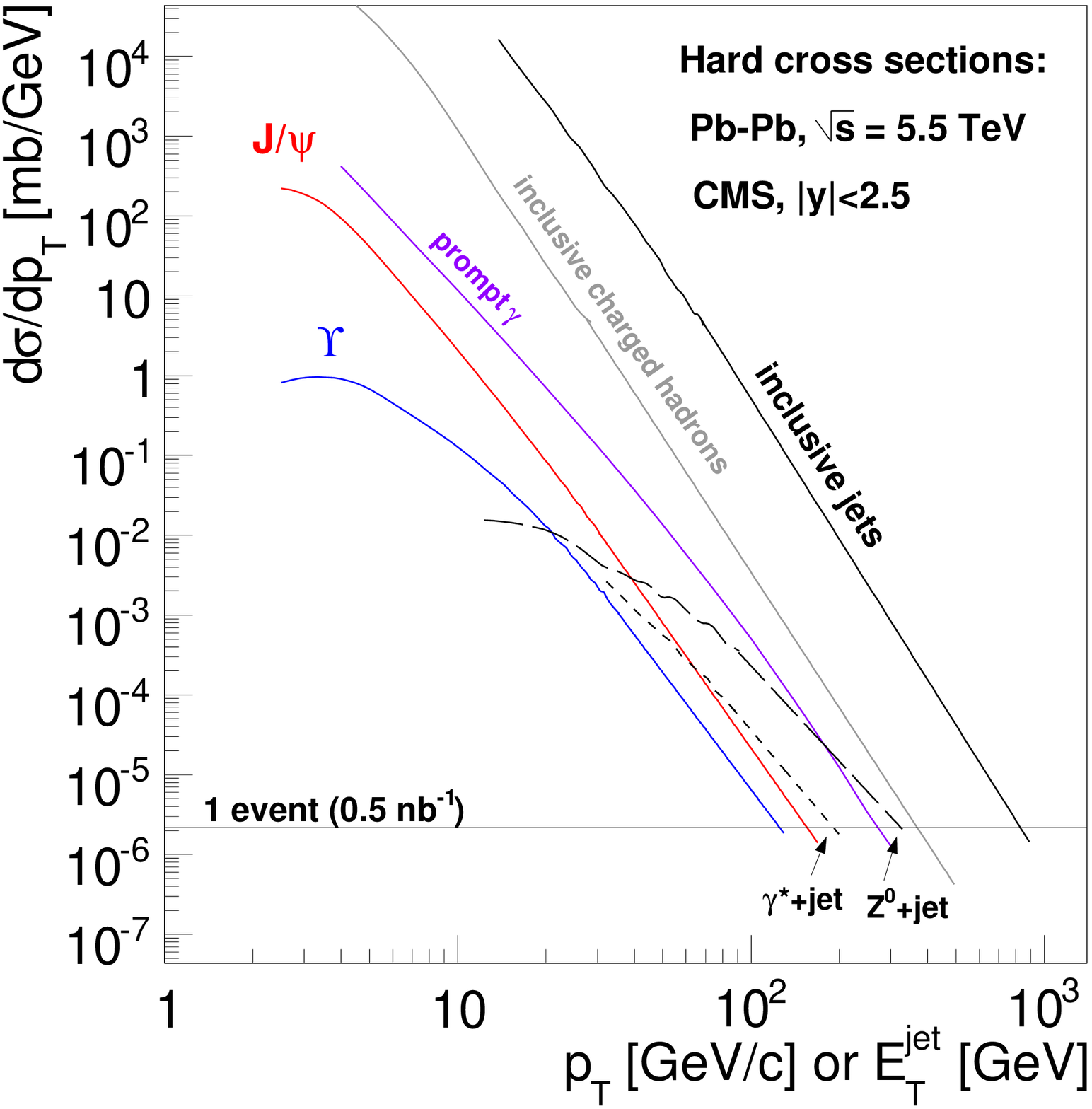}
\caption{Left: Cross sections for various hard processes ($\sigma_{PbPb}^{hard}=A^2\sigma_{pp}^{hard}$) 
in Pb-Pb min. bias collisions in the range $\sqrtsnn\approx$ 0.01--14~TeV.
Right: Differential hard cross sections in Pb-Pb at $\sqrtsnn = 5.5$~TeV in the CMS acceptance~\cite{D'Enterria:2007xr}.}
\label{fig:hard_probes}
\end{figure}

The prime goal of high-energy heavy-ion physics is to study the fundamental theory of the 
strong interaction -- Quantum Chromodynamics (QCD) -- in extreme conditions of temperature, 
density and parton momentum fraction (low-$x$). The collisions of lead nuclei at the LHC 
($\sqrtsnn$~=~5.5~TeV) will probe quark-gluon matter at unprecedented values 
of energy density. A detailed description of the potential of CMS to carry out a series of 
representative Pb-Pb measurements has been presented in~\cite{D'Enterria:2007xr}. 
We summarise this work here plus more recent developments~\cite{Loizides:2008pb}. 
Heavy-ion observables accessible to measurement with CMS include:
\begin{itemize}
\item ``Soft'' observables~\cite{ferenc}: charged hadron multiplicity ($dN_{ch}/d\eta$), 
inclusive identified hadron spectra ($dN/dp_{T}$) and elliptic flow ($v_2$ parameter); 
which provide constraints on the collective properties (entropy density, 
viscosity, ...) 
of the produced strongly interacting medium.
\item Perturbative probes~\cite{Loizides:2008pb,dutta}: quarkonia, heavy-quarks, jets, $\gamma$-jet
and high-$p_{T}$ hadrons; which are produced very early in the collision, are potentially 
modified while traversing the medium, and thus yield ``tomographic'' information (colour density, 
$\qhat$ transport coefficient, critical energy density,  ...) of the hottest and densest phases of the reaction.
\end{itemize}
The LHC will open a new frontier in the study of QCD matter, thanks to
the large energy densities attainable, $\varepsilon_{Bjorken}\approx$~10~GeV/fm$^3$ 
at initial times\footnote{Notice that the ``canonical'' initial-time, $\tau_0$~=~1~fm/c, is 200 times 
larger than the crossing-time of the two nuclei at LHC energies: $\tau_{cross}\approx~2\,R/\gamma\approx~5\,10^{-3}$~fm/c.} 
$\tau_0$~=~1~fm/c; the very low values of parton fractional momenta 
$x\approx p_T/\sqrtsnn \,exp(-\eta)$~=~${\cal O}(10^{-5}$) 
accessible in the collision~\cite{raju}; and the very abundant production 
of hard probes (with cross sections 10 to 10$^4$ larger than at RHIC, see Fig.~\ref{fig:hard_probes}).

%
\section{The CMS experiment}

\begin{figure}[!Hhtb]
  \centering
\includegraphics[width=12.cm,height=6.cm]{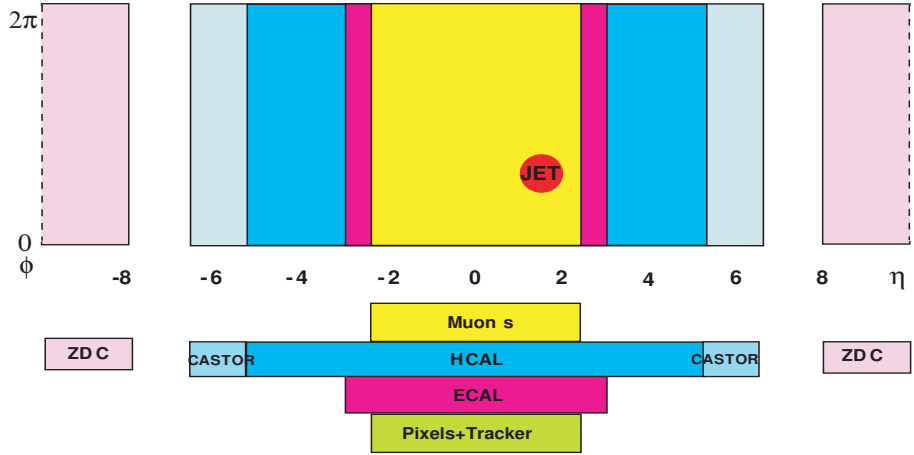}
\caption{CMS coverage for tracking, calorimetry, and muon identification in
pseudo-rapidity ($\eta$) and azimuth ($\phi$). The size of a jet with cone $R=0.5$
is also depicted for comparison.} 
\label{fig:cms_accept}
\end{figure}

\noindent
CMS is a general purpose experiment at the LHC designed to explore the physics at the TeV 
energy scale~\cite{ptdr}. The CMS detector 
is a 22 m (length) $\times$ 15 m (diameter)
apparatus (12\,500 tons in weight) featuring a 4 T solenoid surrounding central silicon 
pixel and microstrip tracking detectors and electromagnetic ($|\eta|<$ 3) and 
hadronic ($|\eta|<$ 5) calorimeters. Muon detectors ($|\eta|<$ 2.4) are embedded in the flux 
return iron yoke of the magnet. CMS is the largest acceptance detector at the LHC
(Fig.~\ref{fig:cms_accept}) with unique detection capabilities in the very forward 
hemisphere with the CASTOR (5.1 $<|\eta|<$ 6.6) and the Zero-Degree (ZDCs, $|\eta_{neut}|>$ 8.3)
calorimeters~\cite{fwd_cms}. 
The detector subsystems have been designed with a resolution and granularity
adapted to cope with the extremely high luminosities expected in the p-p running
mode ($\mathcal{L}\sim 10^{34}$~\percms\ at 14 TeV) with up to 25 simultaneous 
collisions per bunch crossing. The detector can thus perfectly deal with the large particle multiplicities
anticipated for \PbPb\ collisions at 5.5 TeV, where the luminosity is 7 orders of magnitude smaller
($\mathcal{L}\sim 10^{27}$\percms).\\

A key aspect of the CMS capabilities for heavy-ion physics -- in particular to fully exploit the rare 
probes available at the LHC -- is its unparalleled high-level-trigger (HLT) system running on a filter 
farm with an equivalent of ${\cal O}$(10$^4$) 1.8~GHz CPU units, yielding few tens of Tflops~\cite{groland}. 
The HLT system is powerful enough to run  ``offline'' algorithms on {\em every single} Pb-Pb 
event delivered by the level-1 trigger, and select the interesting events while reducing the data stream 
from an average 3~kHz L1 input/output event rate down to 10--100 Hz written to permanent storage. 
The resulting enhanced statistical reach for hard probes is a factor of $\times$20 to $\times$300 larger, 
depending on the signal, than for the min-bias (MB) trigger (Section~\ref{sec:hard}).

%

\section{Soft observables in Pb-Pb collisions at $\sqrtsnn$ = 5.5 TeV}
\label{sec:soft}



\subsection*{(1) Charged hadron Pb-Pb rapidity density ($dN_{ch}/d\eta$): initial entropy (gluon) density}

\begin{figure}[!Hhtb]
\includegraphics[width=7.6cm,height=6.8cm]{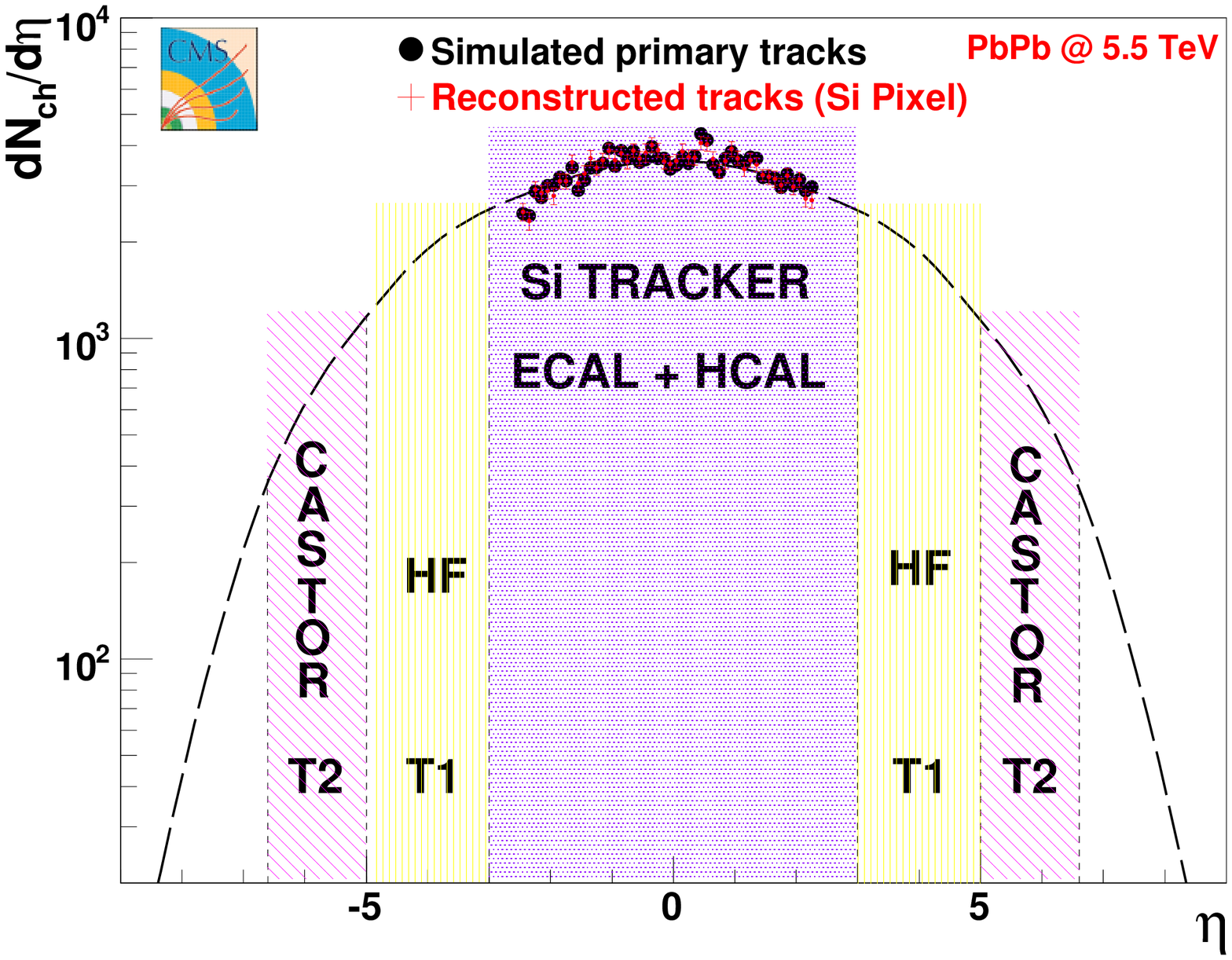}
\includegraphics[width=8.7cm,height=6.75cm]{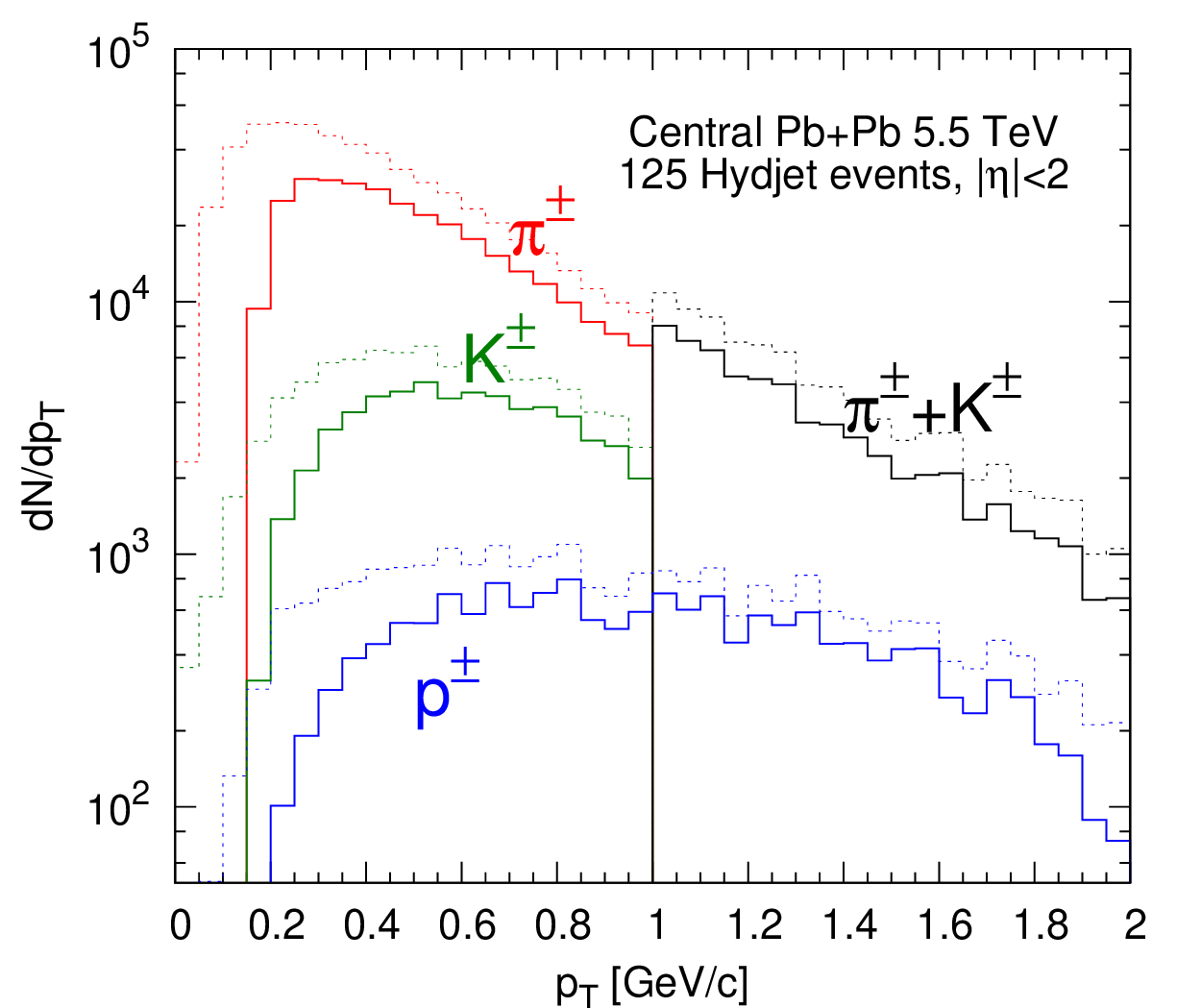}
\vskip -0.4cm
\caption{Bulk hadron production in central Pb-Pb collisions at 5.5 TeV~\protect\cite{ferenc}. 
Left: Pseudo-rapidity distribution of primary tracks simulated with {\sc hijing} (black dots) 
compared to the reconstructed hits in the first layer of the Si tracker (red crosses)
within  $|\eta|<$ 2.5. Right: Reconstructed (solid) and generated (\hydjet, dotted)
$p_{T}$ spectra of pions, kaons and protons.} 
\label{fig:soft}
\end{figure}

\noindent
The charged-particle multiplicity  per unit rapidity, $dN_{ch}/d\eta$, is directly related to the 
entropy density generated in \PbPb\ collisions which in itself fixes the initial global properties of the 
produced matter. Colour Glass Condensate (CGC) approaches~\cite{raju}, which effectively take 
into account a reduced initial parton flux in the colliding nuclei, can reproduce the centrality and 
center-of-mass (c.m.) energy dependences of the bulk A-A hadron production at RHIC.
At the LHC, the relevance of low-$x$ QCD effects will be significantly enhanced since the factor 
of 30 increase in the c.m.~energy, implies a similar reduction in the range of typical fractional 
momenta of the colliding gluons ($x\sim Q/\sqrts$). Current Pb-Pb multiplicities expected at the LHC 
are $dN/d\eta|_{\eta=0}\approx$~1200--2000~\cite{nestor}. Given the excellent granularity
of the Si pixel detector (66M channels in total), a hit counting measurement {\it \a`a la} 
PHOBOS~\cite{phobos} in the innermost pixel layer is possible with very low occupancies (less than 2\%).
Figure~\ref{fig:soft} (left) shows the generated and reconstructed primary hadron multiplicity
within $|\eta|<$ 2.5 in central Pb-Pb collisions.

\subsection*{(2) Soft hadron spectra ($dN/dp_T$): medium Equation-of-State (EoS)}

Measurement of the bulk pion, kaon and (anti)proton spectra in \PbPb\ at 5.5 TeV 
and their comparison to hydrodynamical models will provide the first estimates 
of the conditions characterising the initial- (thermalisation time, 
baryochemical potential) and final- (freeze-out temperature) states, as well as 
the first constraints on the EoS of the produced medium.\\

Despite its large magnetic field, CMS can reconstruct very low $p_T$ tracks down to 
0.2~GeV/c -- by making use of {\it just} the three highly segmented layers of the silicon 
pixel tracker and of its analogue readout -- with a modified hit triplet finding algorithm and 
a (cluster-shape-based) cleaning procedure which reduces the fake rate in high particle 
density environments~\cite{ferenc}. The energy loss of the tracks, $dE/dx$, can be estimated, 
in addition, from the charge deposited in the individual pixels of the clusters, providing 
particle identification in a limited $p_T$ range.
Inclusive yields can be extracted up to $\approx 1$~GeV/c (2~GeV/c) for pions
and kaons (for p, $\bar{\rm p}$) via Gaussian unfolding of the
measured $dE/dx$ distributions. The expected identified hadron 
$p_{T}$ spectra in central Pb-Pb collisions are shown in Fig.~\ref{fig:soft}, right.

\subsection*{(3) Elliptic flow ($v_2$ parameter): medium viscosity}

The observation of strong azimuthal anisotropies with respect to the reaction plane, 
has been one of the highlight results in Au-Au collisions at RHIC. The good agreement 
of the elliptic flow ($v_2$) data with {\it ideal} relativistic hydrodynamics indicates 
that the produced matter develops a strong collective flow in the first fm/c of the collision 
and behaves as a ``perfect fluid'', with a viscosity near a conjectured lower bound~\cite{Kovtun:2004de}. 
The measurement of the differential $v_2$ in \PbPb\ collisions at the LHC will be of primary
importance to constrain the viscosity of the produced matter and test its liquid-like
(or weakly-interacting) properties.\\

The reaction plane can be independently determined with different CMS 
detector subsystems using various analysis methods~\cite{ferenc,gyulnara}. 
At central rapidities ($|\eta|<$ 2.5) CMS will be able to determine the reaction plane 
using the tracker and calorimeters, whereas at forward rapidities $v_2$ will be measured
with the HF and CASTOR calorimeters, in a region almost free from non-flow contributions. 
At beam-rapidity, the ZDCs can provide an independent determination from the directed 
flow signal ($v_1$). The expected precision of the reaction-plane measurement will give 
access to azimuthal anisotropy studies for charged particles (identified or not) in a
momentum range from a few hundred MeV/c up to a few hundred GeV/c.

%

\section{Hard observables in Pb-Pb collisions at $\sqrtsnn$ = 5.5 TeV}
\label{sec:hard}

CMS is, by design, very well adapted to detect (reconstruct) high-$p_T$ (high-mass) particles. This section covers various 
studies that show the capabilities of the detector to measure high-$p_T$ hadrons, full jets, photon-jet processes, and 
quarkonium production in Pb-Pb collisions.

\subsection*{(4) High-$p_T$ hadron and jet spectra ($R_{AA}(p_T)$, $d^2N_{jet}/d\eta dp_T$): medium transport coefficient} 

\begin{figure}[htb]
\hspace{-0.5cm}\includegraphics[width=9cm,height=6.cm]{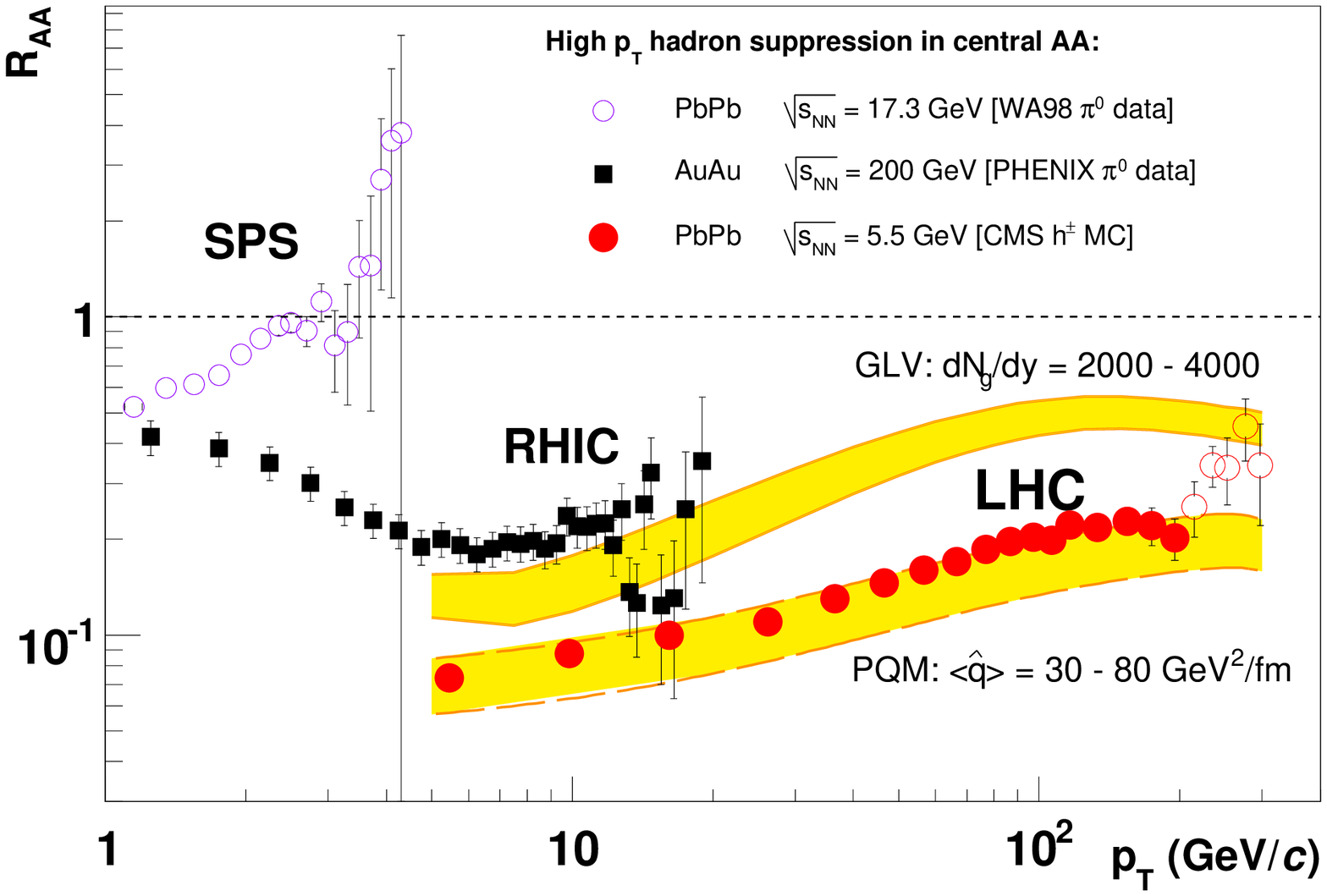}
\includegraphics[width=7.cm,height=6.9cm]{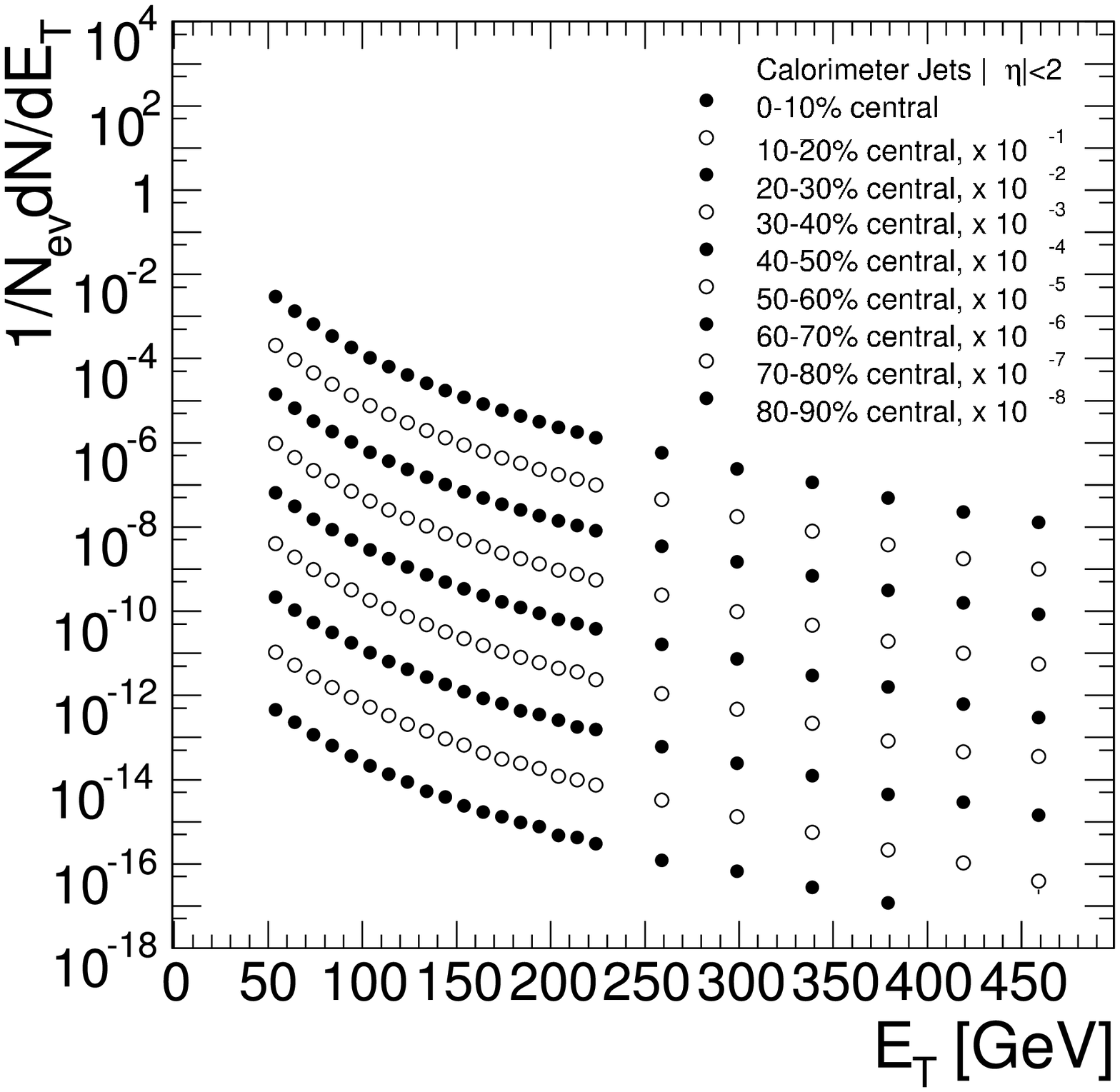}
\caption{Left: Nuclear modification factor, $R_{AA}(p_{T})$, for high-$p_{T}$ hadrons 
at CERN-SPS (open dots) and RHIC (squares) compared to CMS pseudo-data  (closed circles: MB, 
open circles: HLT) and to theoretical LHC expectations from GLV~\protect\cite{Vitev:2002pf} 
($\dNgdy$ = 2000--4000) and PQM~\protect\cite{Loizides:2006cs} ($\qhat\approx$  30--80 GeV$^2$/fm)
models. Right: Inclusive jet $E_{T}$ distributions in 10 centrality bins 
expected in CMS for \PbPb\ at 5.5~TeV (0.5 nb$^{-1}$) using MB- and HLT-triggered data.}
\label{fig:RAA}
\end{figure}

The suppression of inclusive hadron production at high transverse momentum in central Au-Au 
compared to p-p collisions, has been one of the most important observables at RHIC to 
study medium-induced parton energy loss. The measured suppression factor ($R_{AA}$) 
provides information on the initial gluon density, $\dNgdy$, and on the dissipative properties 
of the medium quantified by its transport coefficient, $\qhat$~\cite{CasalderreySolana:2007zz}.
CMS can significantly extend the $p_T$ reach with respect to RHIC, thanks to the large hard 
cross sections at 5.5~TeV, the large acceptance of its tracking system ($|\eta|<2.5$), 
and its high-$p_T$ triggering capabilities. The (complete) silicon tracker has excellent reconstruction performances
attaining a 60\% (total) efficiency above $p_{T}>$~1~GeV/c with a fake track rate below 5\% for central Pb-Pb.
Within $|\eta|<$1, the $p_T$ resolution is better than 2\% up to the highest $p_T$ values reachable.
The leading hadron suppression can thus be measured with low uncertainties all the way up to
300 GeV/c, allowing us to clearly discriminate between various model predictions (Fig.~\ref{fig:RAA}, left).

Full jet reconstruction in Pb-Pb collisions can be performed in CMS using the ECAL and HCAL calorimeters
with an iterative cone algorithm ($R=0.5$ cone size) and subtracting the underlying 
soft background on an event-by-event basis~\cite{D'Enterria:2007xr}. Jets start to be distinguishable 
above the background at $E_T\sim$30 GeV and can be fully reconstructed above 75~GeV
(efficiency and purity close to 100\%) with a good energy resolution (better than 15\%). 
Figure~\ref{fig:RAA} (right) shows the expected jet $E_{T}$ spectrum 
after one month of \PbPb\ running (0.5 nb$^{-1}$), taking into account High Level Trigger
bandwidths and quenching effects as implemented in \hydjet. The large $E_T$ reach, 
up to 0.5 TeV in central \PbPb, will allow us to carry out detailed differential 
studies of jet quenching phenomena (jet shapes, energy-particle flow within the jet, ...).

\subsection*{(5) Photon-tagged jet production: medium-modified fragmentation functions}

The $\gamma$-jet (and $Z$-jet) channel provides a very clean means to determine parton fragmentation 
functions (FFs)~\cite{Arleo:2004xj} which in turn are very sensitive to the underlying $\qhat$ 
parameter~\cite{Armesto:2007dt}. Since the prompt $\gamma$ is not affected by final-state interactions, 
its transverse energy ($\etg$) can be used as a proxy of the away-side parton energy ($\etj\approx\etg$)
{\it before} any jet quenching has taken place in the medium. The FF, i.e. the 
distribution of hadron momenta, $1/{N_{jets}}\,dN/dz$, relative to that of the parent parton $\etj$,
can be constructed using $z = p_{T}/\etg$ or, similarly, $\xi=-\ln z=\ln (\etg/p_{T})$, for all particles with 
momentum $p_T$ associated with the jet.

\begin{figure}[htb]
\includegraphics[width=7.7cm,height=6.5cm]{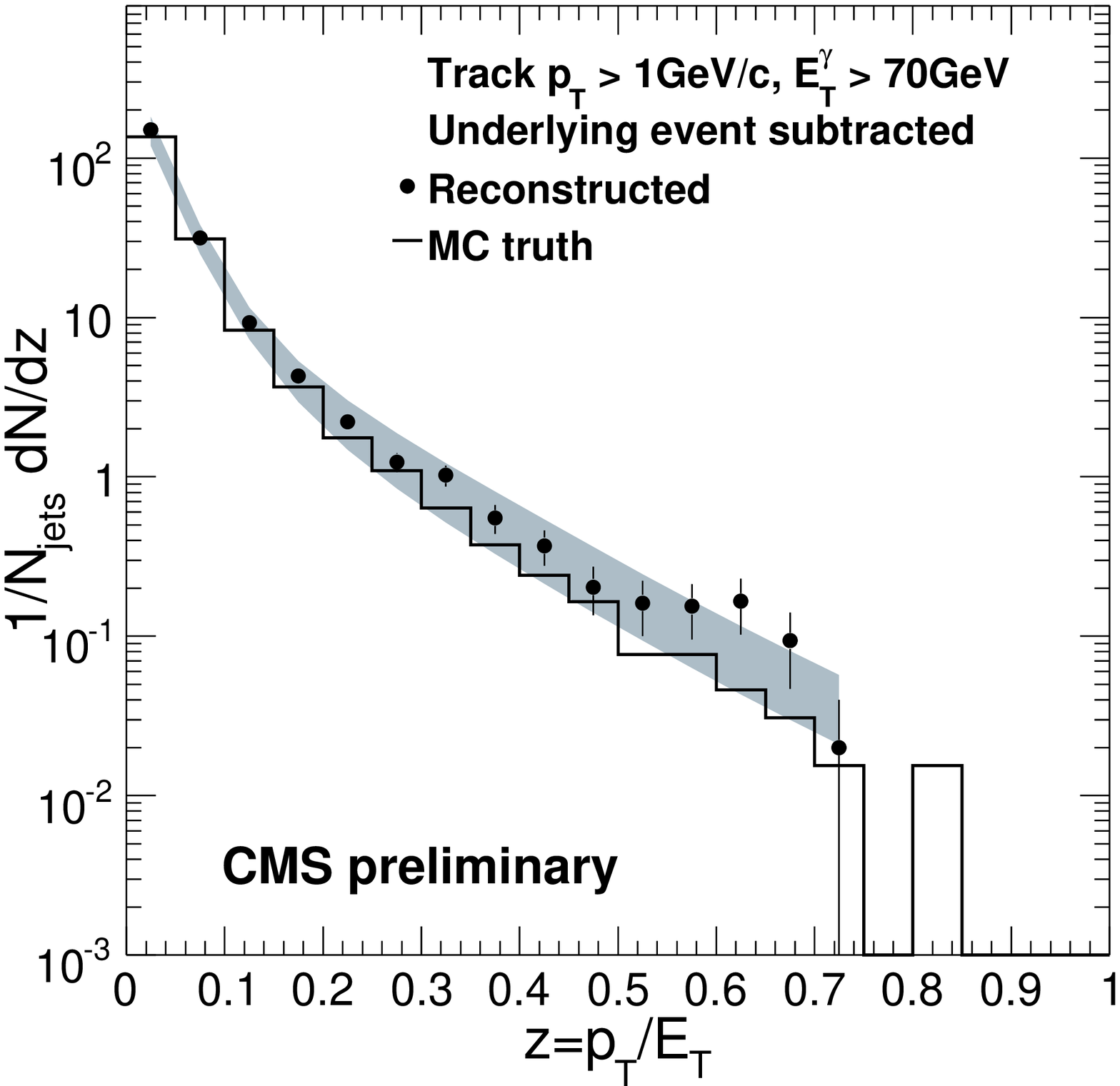}
\includegraphics[width=7.9cm,height=6.8cm]{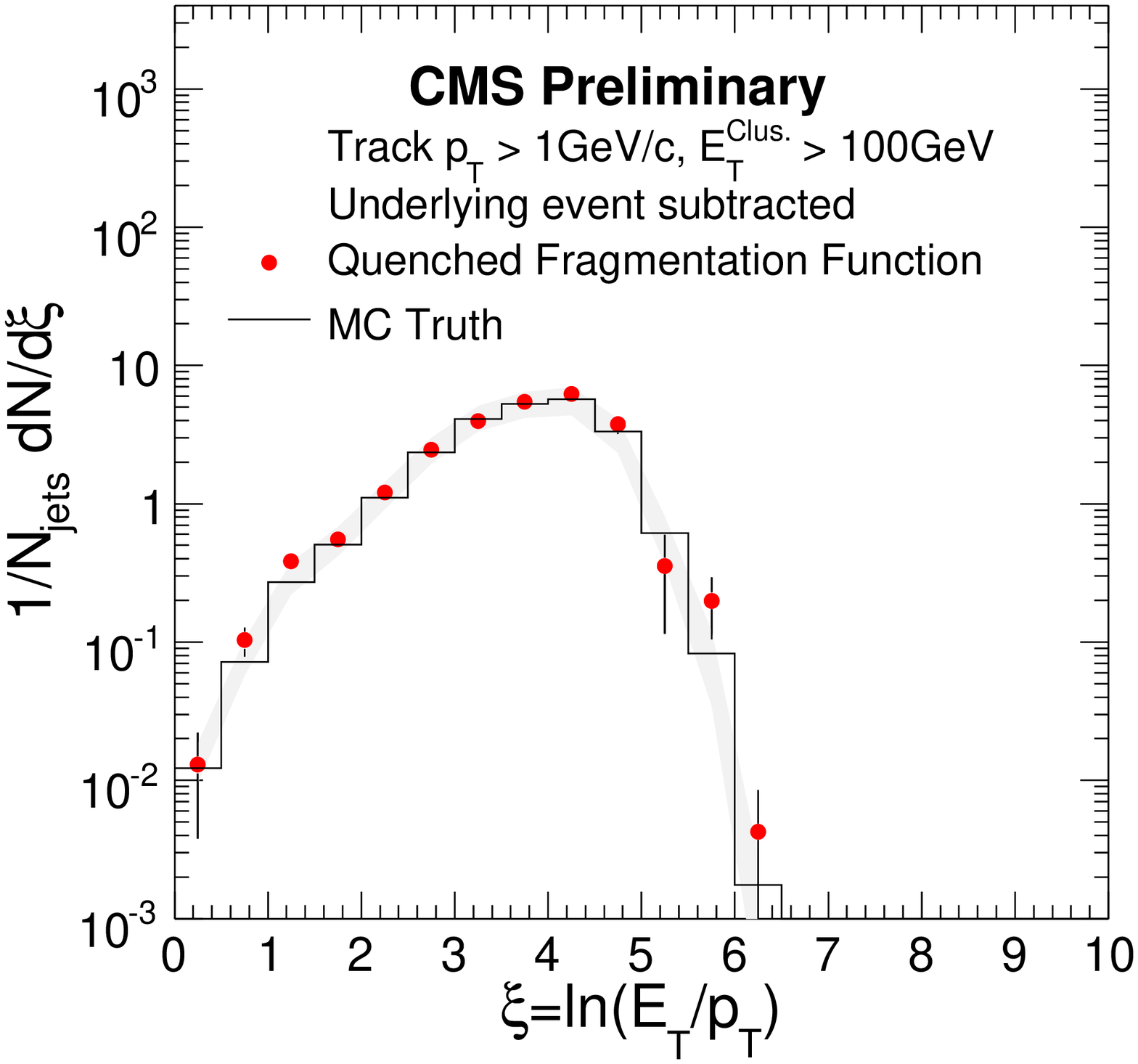}
\caption{Generated (histogram) and reconstructed (points and error band) fragmentation functions 
as a function of $z$ (left) and $\xi$ (right) for quenched partons, obtained in $\gamma-$jet 
events with $\etg>$ 70 GeV in central Pb-Pb at 5.5 TeV (statistical errors correspond to 
0.5~nb$^{-1}$)~\protect\cite{Loizides:2008pb}.}
\label{fig:mFFs}
\end{figure}

Full CMS simulation-reconstruction studies of the $\gamma$-jet channel have been carried out, 
where the isolated $\gamma$  is identified in ECAL ($R_{isol}$~=~0.5), the away-side jet {\it axis} 
($\Delta\phi_{\gamma-jet}>$~3~rad) is reconstructed in ECAL+HCAL, and the momenta of hadrons
around the jet-axis ($R_{jet}<$~0.5) are measured in the tracker~\cite{Loizides:2008pb}. 
A total of 4300 $\gamma$-jet events are expected in one Pb-Pb year at the nominal luminosity 
(according to {\sc pythia} simulations scaled by the Glauber nuclear overlap). 
The obtained FFs for photon-jet events with $\etg>70$~GeV --- after subtraction of the 
underlying-event tracks using a R~=~0.5 cone {\it transverse} to the jet --- are shown in 
Fig.~\ref{fig:mFFs} for central Pb-Pb. Medium modified FFs are measurable with high significance 
(the systematic uncertainties being dominated by the low jet reconstruction efficiency for
$\etj$~=~30--70~GeV) in the ranges $z<$~0.7 or 0.2~$<\xi<$~5.

\subsection*{(6) Quarkonium production: critical energy density }

Studies of quarkonium production in \PbPb\ at the LHC will provide crucial information 
on the thermodynamical state of the QCD medium formed in these reactions. Lattice 
calculations can predict, in principle, the expected step-wise ``melting'' pattern of the 
different $\qqbar$ states due to colour screening as a function of energy density.
The $\ups$, accessible for the first time with large statistics at the LHC, 
is expected to survive up to 4 times the critical QCD temperature and, thus, 
suppression of the $b\bar{b}$ ground state would be indicative of medium 
temperatures around 1 GeV.

\begin{figure}[htb]
\hspace{-1cm}
\includegraphics[width=9cm,height=6.5cm]{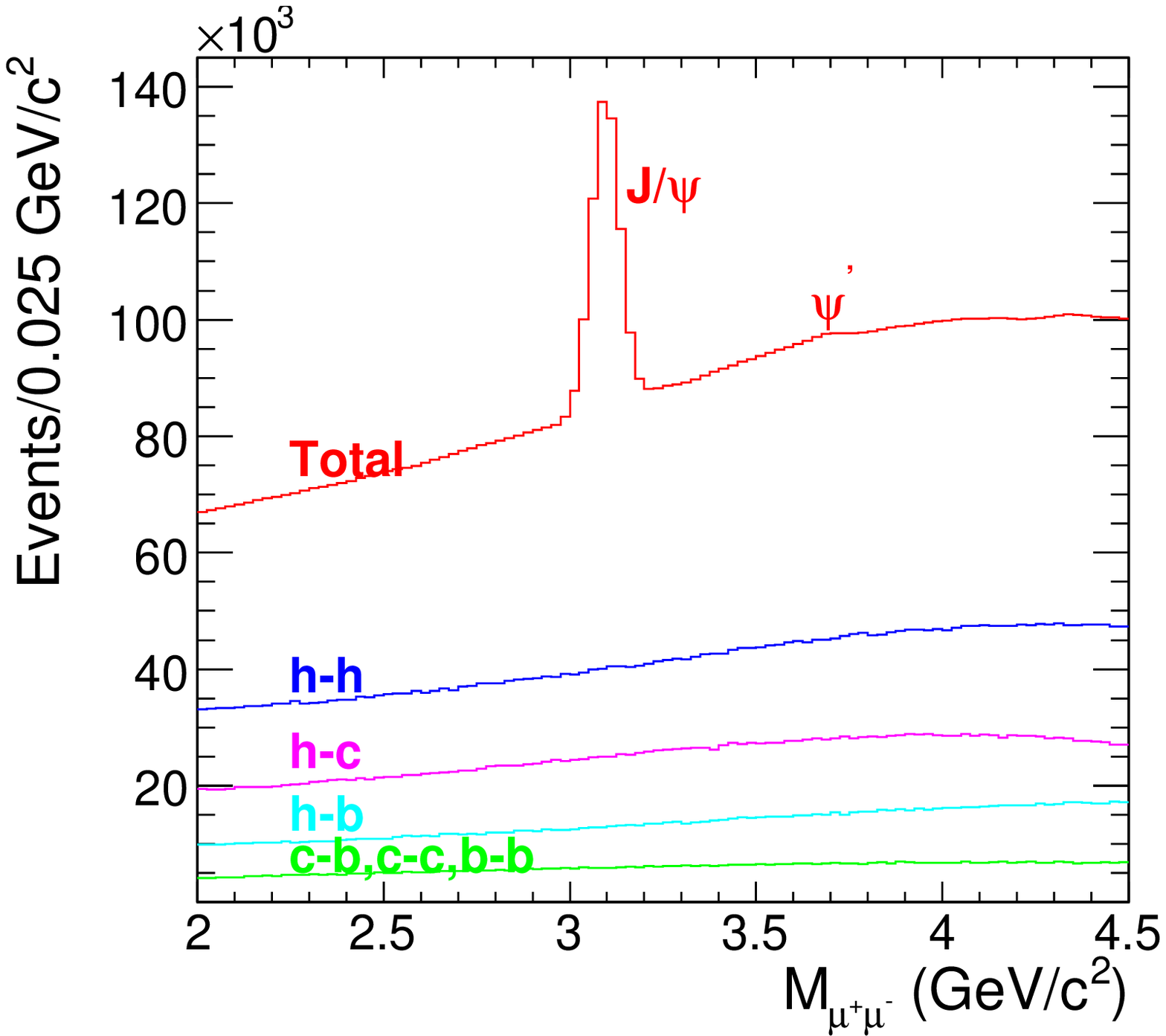}
\hspace{-0.9cm}\includegraphics[width=9cm,height=6.5cm]{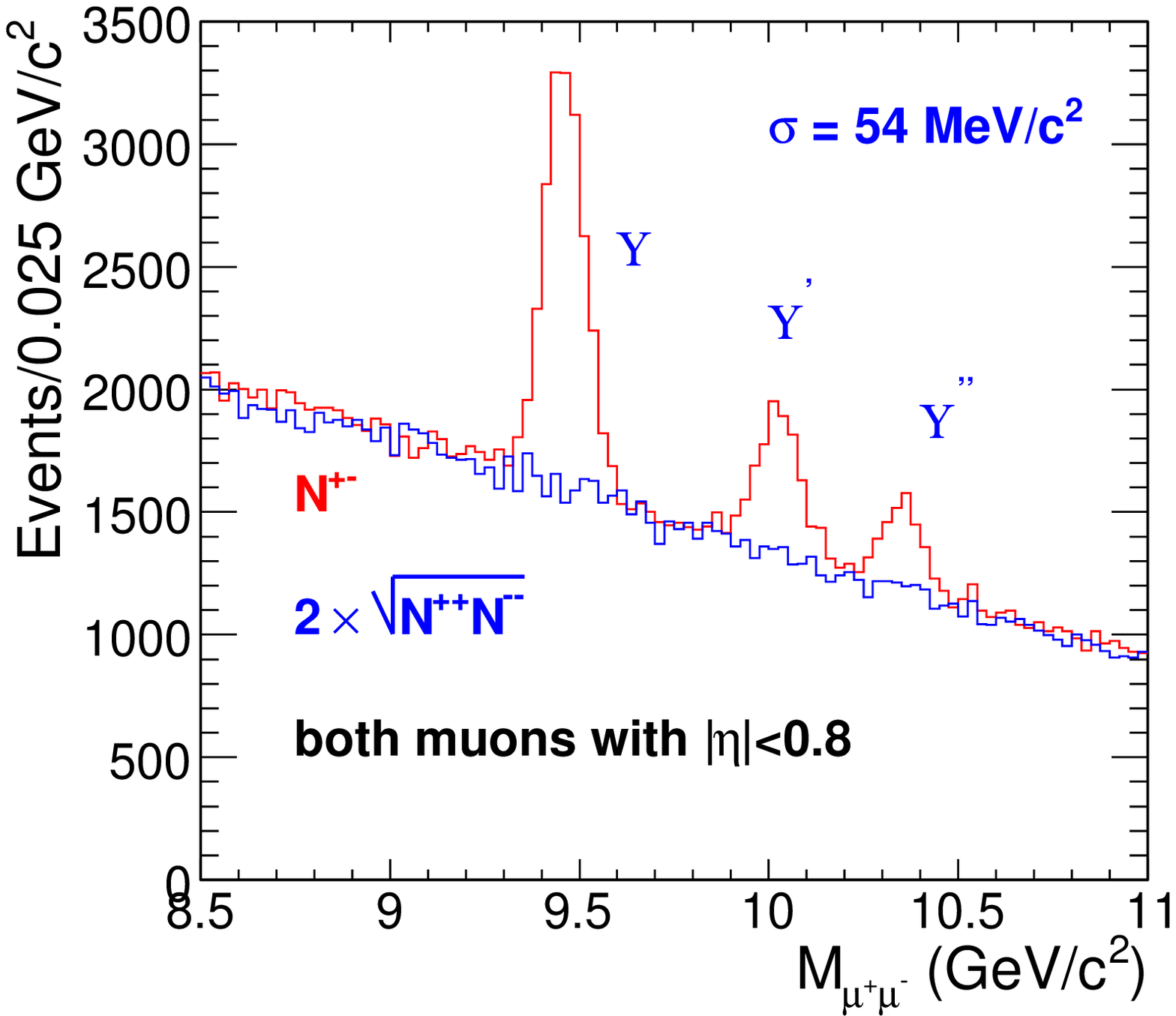}
\caption{Full simulation studies of quarkonia production in Pb-Pb at 5.5 TeV ($\dNdeta$~=~2500). 
Left: Dimuon mass distributions within $|\eta|<$ 2.4 (the $h$, $c$ and $b$ backgrounds 
stand for decay muons from $\pi+K$, charm, and bottom, resp.). 
Right: Invariant mass spectra of opposite-sign and like-sign muon pairs (within $|\eta|<0.8$)
in the $\Upsilon$ mass region.}
\label{fig:qqbar}
\end{figure}

CMS can reconstruct the $c\bar{c}$ and $b\bar{b}$ resonances via their dimuon 
decay channel with  
very high efficiencies ($\sim$\,80\%) and purities ($\sim$\,90\%), and the 
best LHC mass resolution: 54~MeV/c$^2$ at the $\ups$ mass, in the central barrel ($|\eta|<0.8$).
The $\ups$ states can be measured all the way down to $p_{T}$ =~0~GeV/c with 
acceptances between 15\% and 40\%. 
The large aperture of the muon detectors and the precise tracking result in
a very good separation between the $\qqbar$ states in the dimuon mass distributions,
with high statistics and good signal to background ratios (Fig.~\ref{fig:qqbar}).
In the absence of suppression, after one month of \PbPb\ running (0.5~nb$^{-1}$) 
CMS will collect 1.8\,10$^5$ $\jpsi$ and 2.5\,10$^4$ $\ups$, enough to compare 
central and peripheral \PbPb\ collisions, and to carry out differential studies 
($dN/dy$, $dN/dp_{T}$) that will help to clarify the mechanisms behind the production 
(and suppression) of  $\qqbar$ states in high-energy nucleus-nucleus collisions.

\subsection*{(7) Exclusive $\ups$ photoproduction: low-$x$ parton distribution functions (PDFs)}

\begin{figure}[htb]
\includegraphics[width=7.7cm,height=5.8cm]{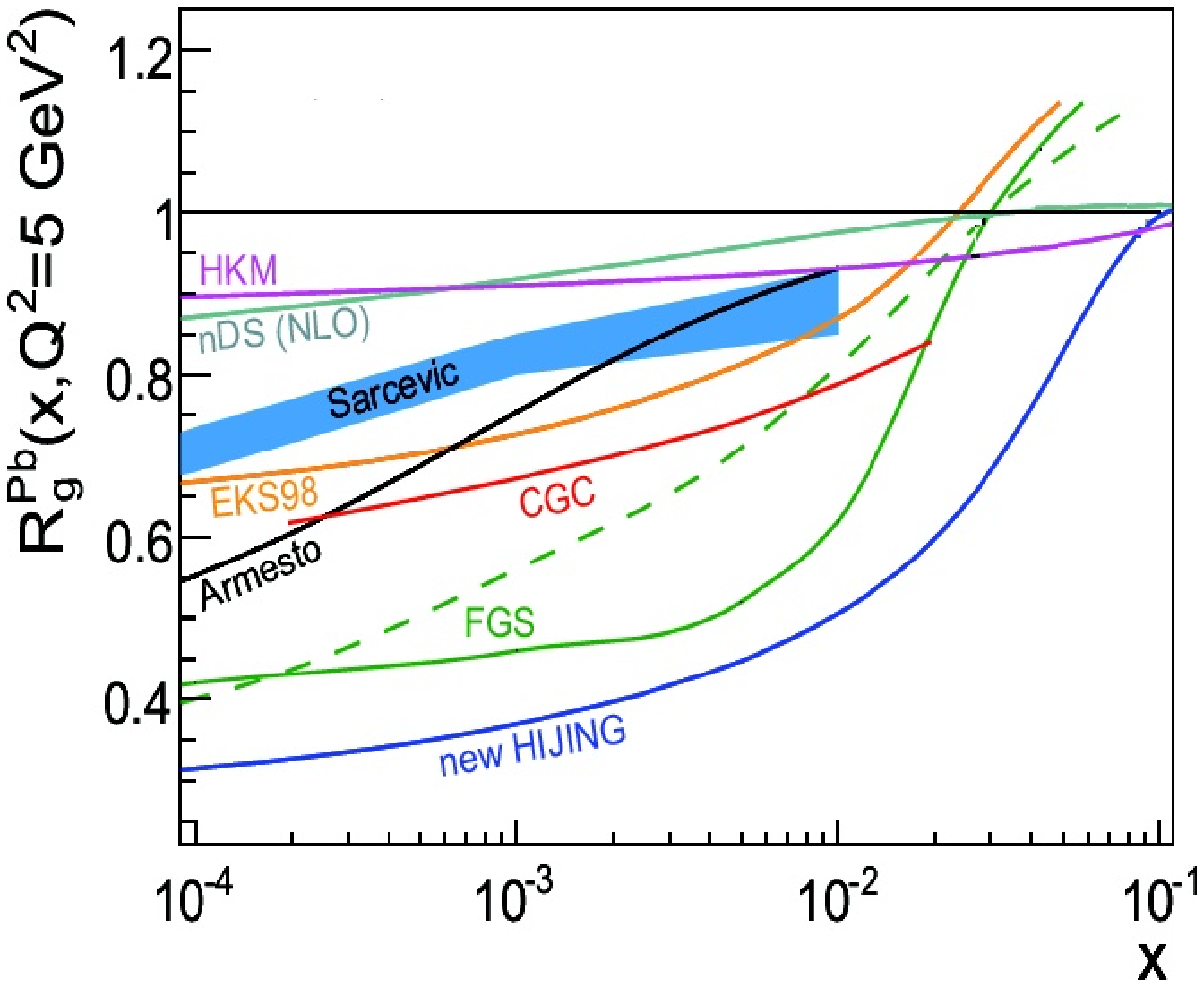}
\includegraphics[width=7.7cm,height=6.cm]{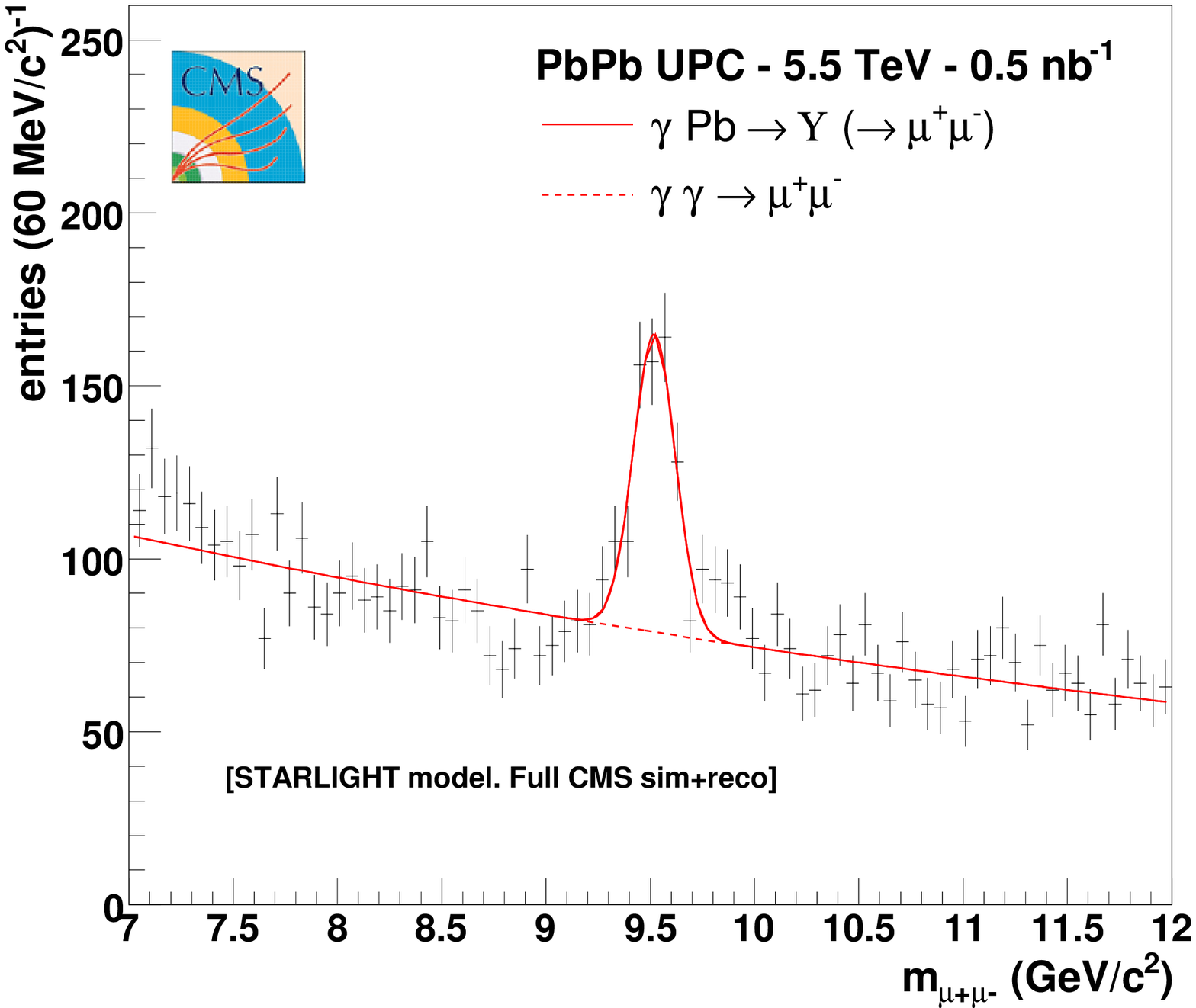}
\caption{Left:  Pb-over-proton gluon densities ratio as a function of $x$ at fixed $Q^2$~=~5~GeV$^2$ 
from various PDF fits. 
Right: Expected $\mumu$ invariant mass distribution from $\gamma\,$Pb$\rightarrow \Upsilon\,$Pb$^\star$ 
($\ups\rightarrow \mumu$, signal) and $\gaga\rightarrow \mumu$ (background) in UPCs triggered 
with the ZDC.}
\label{fig:upc_ups_cms}
\end{figure}

\noindent
The strong electromagnetic fields (equivalent fluxes of quasi-real photons) in 
ultraperipheral heavy-ion collisions (UPCs), where both nuclei pass at distances 
$R\gtrsim$~20~fm, can be used to study various $\gamma\gamma$ and 
$\gamma$-nucleus processes up to c.m.~energies 
$W_{\gaga}^{\ensuremath{\it max}}\approx$~160~GeV and 
$W^{\ensuremath{\it max}}_{\gA}\approx$~1~TeV \cite{Baltz:2007kq}. 
The Bjorken-$x$ values probed in quarkonium photoproduction processes 
($\gpb\rightarrow\ups \,$Pb) can be as low as $x\sim 10^{-4}$ where the nuclear PDFs, 
in particular the gluon one, have very large uncertainties (Fig.~\ref{fig:upc_ups_cms}, left).

Full simulation+reconstruction studies show that CMS can measure $\ups\rightarrow \elel$, $\mumu$ 
within $|\eta|<$ 2.5, in UPCs tagged with neutrons detected in the ZDCs. Figure~\ref{fig:upc_ups_cms} (right)
shows the reconstructed $dN/dm_{\mumu}$ around the $\ups$ mass. The expected total yield of 
$\sim$\,400~$\ups$/year  in the dielectron and dimuon channels will allow us to carry out 
differential ($p_T$,$\eta$) studies to help constrain the low-$x$ gluon density in the Pb nucleus.

\noindent
{\bf Acknowledgments.} Supported by 6th EU Framework Programme MEIF-CT-2005-025073.\\ 
\vspace*{-.8cm}

\section*{References}

\end{document}